\documentclass[12pt]{iopart}
\usepackage{epsf} 
\begin{document}

\title[Ising model on the Sierpinski carpet fractal]{Spontaneous magnetization 
of the Ising model on the Sierpinski carpet fractal, a rigorous result}
\author{A. Vezzani 
\footnote{vezzani@fis.unipr.it}} 
 
\address{Istituto Nazionale Fisica della Materia (INFM).\\ 
Dipartimento di Fisica, Universit\`a di Parma, parco Area delle Scienze 7A 
43100 Parma Italy}

\begin{abstract}
We give a rigorous proof of the existence of spontaneous magnetization at
finite temperature for the Ising spin model defined on the Sierpinski
carpet fractal. The theorem is inspired by the classical Peierls argument
for the two dimensional lattice. Therefore, this exact result proves the 
existence  of spontaneous magnetization for the Ising model in low dimensional 
structures, i.e. structures with dimension smaller than 2. 
\end{abstract}


\submitto{\JPA}


\section{Introduction}

The study of the physical models on the Sierpinski carpet \cite{mandel} has 
been the focus of several works in the last two decades. However, most of these 
papers are based on approximate renormalization group techniques
\cite{rgising,rgcarpet}, on perturbative expansions \cite{htexpising} or
on numerical simulations \cite{numising,numcarpet}. Thus, very few 
rigorous results are known for this self-similar structure. Indeed,
the Sierpinski carpet is an infinitely ramified fractal, where it is not
possible to apply the exact decimation techniques, which allow to solve many
models, defined on finitely ramified fractals, such as
the Sierpinski gasket \cite{gasketising,gasket}. 

A relevant theorem is given by an upper and a lower bound 
on the return probability of a simple random walk \cite{barlow}. 
These bounds prove that the spectral dimension \cite{specdim}
of the Sierpinski carpet is smaller than two. Hence,
from the generalized Mermin-Wagner theorem \cite{mwgen}, the 
continuous symmetry spin models do not present 
spontaneous magnetization. On the other hand, there are no general criteria 
relating the behaviour of the discrete symmetry models, in particular Ising, 
with a simple parameter, describing the topology of the space where the model 
is defined. The main known
results regard the euclidean lattices, where spontaneous magnetization is 
present if the dimension is $\geq 2$, and the exactly decimable fractals, which
never present spontaneous magnetization \cite{gasketising}.

However, the approximate calculations \cite{rgising,htexpising} and the
numerical simulations \cite{numising} suggest that an Ising spin 
system on the Sierpinski carpet is spontaneously magnetized at finite
temperature. Recently, this property has been proved in \cite{shinoda}, 
where it is shown that it follows from the
behaviour of percolation on the Sierpinski carpet. Here we give an
alternative proof inspired by the classical Peierls argument \cite{peierls} for 
the Ising model on the two-dimensional lattice and, in particular, by the 
version given in \cite{peierls2}. These theorems are the first exact results 
showing the existence  of Ising transition
in a low dimensional structure, i.e. in a structure with spectral 
dimension smaller than 2.

\begin{figure}
\begin{center}
\leavevmode
\hbox{%
\epsfysize=6cm
\epsfbox{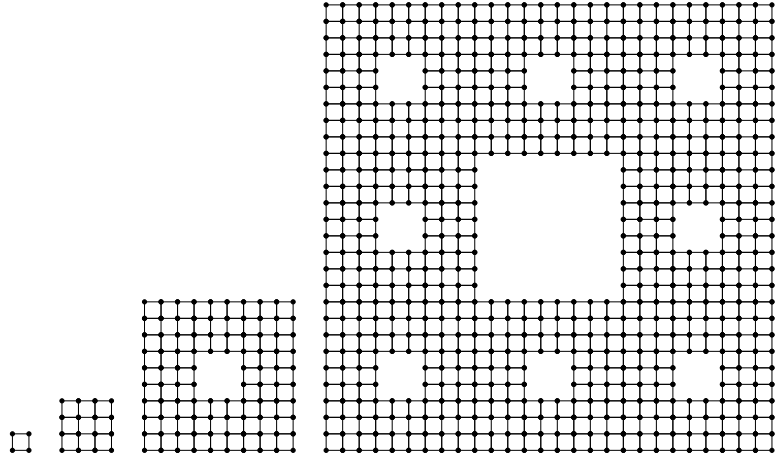}}
\caption{The first 4 generations of the Sierpinski carpet graph. The sites of
the graph (denoted with points) represent the spins and the links (denoted with
lines) represent the interactions.}
\end{center}
\end{figure}

We will consider the Sierpinski carpet graph illustrated 
in Figure 1, even if the proof can be generalized to different infinitely 
ramified carpets. To each site of the graph we associate an Ising spin and to 
each link a ferromagnetic interaction. We call $p_i^-$ the probability for 
the magnetization of the spin $i$ to be smaller than one and we show 
that, at low enough temperature,  $p_i^-< \epsilon <1/2$ for each site $i$. 
The theorem is based on a low
temperature loop expansion of this probability and on an estimate of the number
of the self-avoiding on the sites (SOS) loops. 

The main difference from the two-dimensional case is that
now the space, called dual graph, where the
loops are defined, has unbounded connectivity, namely, the coordination 
(number of nearest neighbours) of a site is finite but not bounded. Hence the 
proof requires more complicated mathematical techniques. Notice that the
standard Peierls argument results rather simple, as the dual of the 
two-dimensional lattice is again a two-dimensional lattice. In spite of the 
unbounded connectivity, here we will be able to estimate the number
of SOS loops because of an important property of the dual graph, which will be 
called {\it sparse high connectivity}, i.e. sites with large 
coordination are far apart. In particular, when the coordination 
is greater than $4\cdot 3^m$, the sites are distant at least $2^{m+1}$
in the intrinsic metric of the dual graph.

Peierls argument is valuable because of the insight it gives into why the Ising 
model is magnetized. Hence, we believe that this proof 
could be useful to find a general criterion for the existence of spontaneous
magnetization on a generic discrete structure. Notice that Peierls argument has 
been already used to show the existence of Ising transition in an inhomogeneous
systems, in particular in the case of diluted random graphs above the 
percolation tresholds \cite{chayes}. 
We start introducing some definitions and notations.

\section{Definitions and notations}

Let us consider the $N$ generation of a Sierpinski carpet. On each site $i$ 
we define an Ising spin $s_i=\pm 1$ and we fix the spins on
the boundary to the value $+1$ (see Figure 2). The energy of the spin system is 
given by:
\begin{equation}
\label{eq1}
-{1 \over 2}\sum_{\langle ij\rangle} s_i s_j
\end{equation}
where the sum runs over all the pairs $\langle ij\rangle$ of nearest neighbour
sites, i.e. sites connected by a link.
In the canonical ensemble the probability for the magnetization of the spin
$i$ to be smaller than $1$ is:
\begin{equation}
\label{eq2}
p_i^- = {\displaystyle \sum_{\{s_1,\dots s_N \}}^{s_i=-1} 
e^{\beta /2 \sum_{\langle ij\rangle} s_i s_j} \over \displaystyle
\sum_{\{s_1,\dots s_N \}}
e^{\beta /2 \sum_{\langle ij\rangle} s_i s_j}}
\end{equation}
In the denominator we sum over all possible spin configurations with $+1$
boundary conditions. In the numerator over all configurations with $+1$ 
boundary conditions and the spin $i$ fixed to $-1$. 
We will prove the spontaneous magnetization of the
system by showing the existence of a temperature $\bar{\beta}$ such that, in 
the thermodynamic limit, $\forall \beta > \bar{\beta}$ $p_i^- < \epsilon < 1/2$.

\begin{figure}
\begin{center}
\leavevmode
\hbox{%
\epsfysize=6cm
\epsfbox{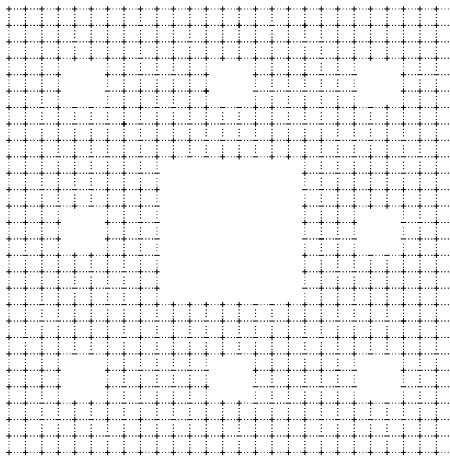}}
\caption{A spin configuration of the Ising spin model defined on the fourth
generation of the Sierpinski carpet. The spins of the boundary are fixed to be 
$+1$. With the dotted lines we represent the interactions.}
\end{center}
\end{figure}

The theorem will be proved in the following steps. In Section III we
introduce the dual space and the loop expansion of equation (\ref{eq2}).  
With this expansion, we prove that $p_i^-$ can be
overestimated by the sum of $\exp(-\beta |l|)$ over all SOS 
loops enclosing $i$ ($|l|$ is the length of the loop $l$). 

In section IV we introduce a tree representation of the SOS loops of length
$|l|$ enclosing 
$i$ and passing the site $\alpha$. This representation allows us to 
overestimate the number of the loops as $\tilde{Z}_{\bar{l}_i^{\alpha}} 
\prod_{\alpha \in \bar{l'}_i^{\alpha}}\tilde{z}_{\alpha}$.  
$\bar{l}_i^{\alpha}$ is a SOS loop. The product runs over 
the sites belonging to 
$\bar{l}_i^{\alpha}$ excluding the site of highest coordination. 
$\tilde{Z}_{\bar{l}_i^{\alpha}}$ is the coordination number
of the site representing, in the tree, the site of highest coordination in 
$\bar{l}_i^{\alpha}$. 

In section V we define the sparse high connectivity property 
and, using this property, we 
show that $\prod_{\alpha \in \bar{l'}_i^{\alpha}}\tilde{z}_{\alpha}\leq 
\exp(K|l|)$ where $K$ is a suitable real constant.

In section VI we prove that $\tilde{Z}_{\bar{l}_i^{\alpha}}\leq 
10|l|^{\log_2(3)}$. In particular
we exploit the fact that $\tilde{Z}_{\bar{l}_i^{\alpha}}$ is smaller
than the number of the nearest neighbours, of the site of highest coordination,
which can be crossed by a loop of length $|l|$ and surrounding $i$. 

In section VII we show that the number of sites, which can be crossed 
by a SOS loop of length $|l|$ and surrounding $i$, is smaller than 
$\pi (8 |l|)^{2\log_2(3)}$. This last estimate is proved by 
evaluating the maximum euclidean 
distance that can be covered in the dual lattice by a loop of $|l|$ steps.
Finally we will obtain 
\begin{equation}
p_i^-\leq \sum_{|l|=0}^{\infty} C |l|^{3\log_2(3)} e^{(K-\beta)|l|}
\label{eq0}
\end{equation}
where $C$ is a suitable positive constant. From equation (\ref{eq0}) we have
that $p_i^- < \epsilon < 1/2$, for small enough temperature. 

Beside the sparse high coordination property, in
the proof we strongly use the fact that the dual lattice is embedded in a two
dimensional euclidean space and the fact that the plaquettes of the carpet are
convex sets of this space.

\section{Loop expansion}

In analogy with Peierls argument, we write
the Hamiltonian of the system in terms of a loop expansion. These loops belong 
to the dual graph of the discrete structure where the model is defined. In
particular, the dual graph of the Sierpinski carpet is obtained by 
introducing a site $\alpha$
for each plaquette of the carpet. Then, two sites are connected 
if and only if it
exists a link of the carpet separating the plaquettes relevant to the sites we
are considering. Hence, there is a one to one correspondence between the links
of the carpet and the links of the dual.
We will call $V_d$ the set of the sites of the dual graph and 
$E_d$ the set of the links. Notice that the dual graph has unbounded
connectivity; hence, it must be studied with a certain caution. 
The length (number of links) of the shortest walk connecting two sites 
$\in V_d$ will be called the chemical distance between them.
In Figure 3 we illustrate the Sierpinski carpet and its dual graph. 

\begin{figure}
\begin{center}
\leavevmode
\hbox{%
\epsfysize=6cm
\epsfbox{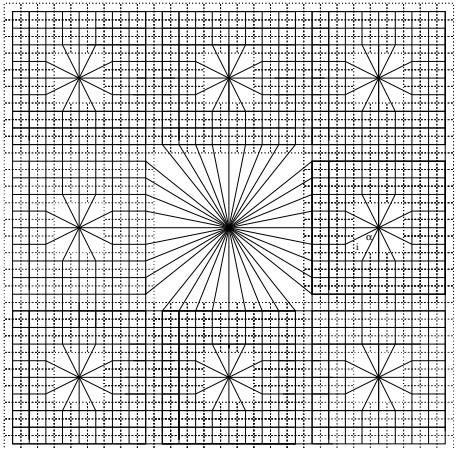}}
\caption{The dashed line denotes the Sierpinski carpet and the continuous line 
the dual graph. The site $i$ belongs to the carpet, and $\alpha$ to the dual
graph.}
\end{center}
\end{figure}

Let us introduce a one-to-one correspondence between the spin 
configurations and the self-avoiding loops of the dual graph (self-avoiding on
the links). A spin configuration with $+1$ boundary conditions is 
identified by the links of the carpet which separate the spins with $+1$ 
magnetization and the spins with $-1$ magnetization. Since, by definition, each 
link of the dual graph corresponds exactly to a link of the carpet, we obtain a 
relation between the spin configuration and
a subset $L \subseteq E_d$. It 
is easy to show that the number of links in $L$
adjacent to the same site $\alpha$ is even. Therefore,
$L$ can always be obtained as a superposition of
self-avoiding (on the links) loops. If in $L$
there are more than two links adjacent to the same site, the decomposition 
of $L$ into loops is not unique. 

On the other hand, let us call $L$ any subset $\subseteq E_d$ such that for any 
$\alpha \in V_d$ the number of links in $L$ adjacent to $\alpha$ is even, 
i.e. $L$ can be decomposed into loops (we will call $\bar{E}_L$ the set of all
possible $L$). In a unique way we can associate to $L$ a 
spin configuration with $+1$ boundary conditions, by fixing to $+1$ the spins
enclosed by an even number of loops and to $-1$ the spins enclosed by an odd
number of loops. In Figure 4 we illustrate a spin configuration and its
corresponding $L\subseteq {E}_d$. 

\begin{figure}
\begin{center}
\leavevmode
\hbox{%
\epsfysize=6cm
\epsfbox{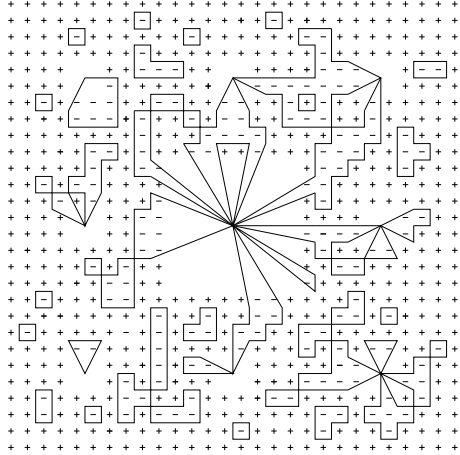}}
\caption{A spin configuration on the Sierpinski carpet
and the corresponding link set $L\subset E_d$.}
\end{center}
\end{figure}

The energy of the spin configuration can be written in terms of the
corresponding set of links $L$. Indeed
\begin{equation}
\label{eq3}
H=-{1 \over 2} \sum_{\langle ij\rangle} s_i s_j=
-|E_d|+{1 \over 4} \sum_{\langle ij\rangle} (s_i -s_j)^2
\end{equation} 
where $|E_d|$ is the total number of links of the dual graph. A link
gives a contribution $1$ to the Hamiltonian (\ref{eq3}), if it connects 
opposite spins, and zero, if it connects parallel spins. Dropping the 
irrelevant additive constant 
$|E_d|$, we obtain $H=|L|$, where $|L|$ is the number of links in the set
$L\in \bar{E}_L$ corresponding to the spin configuration of energy $H$. If 
we consider a decomposition of $L$ into loops $l$ we obtain
\begin{equation}
\label{eq4}
H=|L|=\sum_{l} |l|
\end{equation} 
where $|l|$ is the length of the loop $l$. The probability $p_i^-$ is 
\begin{equation}
\label{eq5}
p_i^- = {\displaystyle \sum_{L^-\in\bar{E}_L^-} 
e^{-\beta |L^-|} \over \displaystyle
\sum_{L\in \bar{E}_L}
e^{-\beta |L|}}
\end{equation}
In the denominator we sum over all possible sets $L\in \bar{E}_L$ and in the 
numerator over all subsets $L^-\in \bar{E}_L$ such that an odd number of loops 
encloses the site $i$. $\bar{E}_L^-$ denotes the set of all $L^-\in\bar{E}_L$, 
such that an odd number of loops encloses $i$; while $\bar{E}_L^+$ denotes the 
set of all $L^+\in\bar{E}_L$, such that an even number of loops encloses $i$. 
Hence, $\bar{E}_L=\bar{E}_L^+\cup \bar{E}_L^-$.

We will call $l_i$ the shortest SOS loop belonging 
to $L^-$ and surrounding $i$ (see 
Figure 5). Each set $L^-$ can be considered as the union of $l_i$ and a suitable
set of $\bar{E}_L^+$. Therefore, equation (\ref{eq5}) can be written as
\begin{equation}
\label{eq6}
p_i^- = {\displaystyle \sum_{l_i} e^{-\beta |l_i|} {\sum_{L^+\in\bar{E}_L^+}}' 
e^{-\beta |L^+|} \over \displaystyle
\sum_{L^+\in \bar{E}_L^+} e^{-\beta |L^+|}+
\sum_{L^-\in \bar{E}_L^-} e^{-\beta |L^-|}}
\end{equation}
$\sum_{l_i}$ denotes the sum over all possible SOS
loops $l_i$ enclosing the site $i$. $\sum_{L^+\in\bar{E}_L^+}'$ is the sum 
over all $L^+\in\bar{E}_L^+$ such that $L^+\cap l_i=\emptyset$ and such that,
in $l_i\cup L^+$, $l_i$ is the shortest SOS loop surrounding $i$. 
Dropping the restriction on the sum 
$\sum_{L^+\in\bar{E}_L^+}'$ we obtain an overestimate of $p_i^-$
\begin{equation}
\label{eq7}
p_i^- \leq {\displaystyle \sum_{l_i} e^{-\beta |l_i|} \sum_{L^+\in\bar{E}_L^+} 
e^{-\beta |L^+|} \over \displaystyle
\sum_{L^+\in \bar{E}_L^+} e^{-\beta |L^+|}+
\sum_{L^-\in \bar{E}_L^-} e^{-\beta |L^-|}}\leq \sum_{l_i} e^{-\beta |l_i|}
\leq \sum_{k=0}^{\infty} n_i(k )e^{-\beta k}
\end{equation}
$n_i(k)$ represents the number of SOS loops of
length $k$ enclosing $i$. Eventually we can write
\begin{equation}
\label{eq8}
p_i^- \leq \sum_{\alpha} \sum_{k=0}^{\infty} n_i^{\alpha}(k) e^{-\beta k}
\end{equation} 
where $n_{i}^{\alpha}(k)$ is the number of SOS loops of length $k$, enclosing 
$i$ and passing the site $\alpha$ of the dual graph. 

\begin{figure}
\begin{center}
\leavevmode
\hbox{%
\epsfysize=5cm
\epsfbox{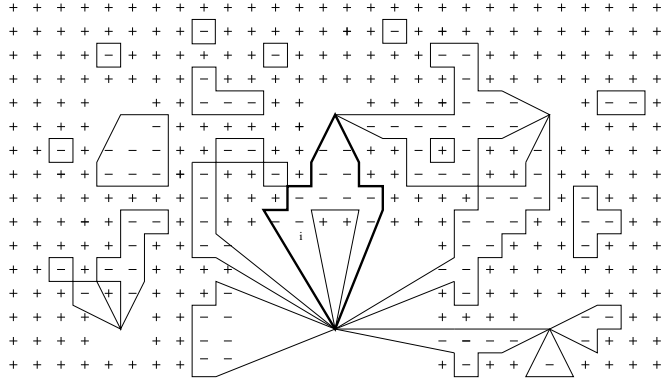}}
\caption{The heavy line denotes the shortest self-avoiding on the sites 
loop in $L$ surrounding the site $i$.}
\end{center}
\end{figure}

Now we take the thermodynamic limit. The theorem
will be proved by showing on $n_{i}^{\alpha}(k)$ an exponential bound
independent from $i$. 
Up to now we followed step by step the Peierls argument for the Ising model on 
two-dimensional lattices \cite{peierls2}. However,
here the dual graph is more intricate than in the two-dimensional case, since 
it has unbounded coordination. Hence, the proof of the bound on 
$n_{i}^{\alpha}(k)$ requires more work.

\section{Tree representation of the loops}

In order to count the SOS loops of length $k$, surrounding $i$ and starting from 
$\alpha$, we represent them into a tree. 
By induction on $N$ we prove that any set of $N$ SOS loops of length 
$k$, starting from $\alpha'$, can be represented in a 
tree satisfying the following properties (TP).

Each site of the tree $a$ is associated to a site of the dual graph $\alpha(a)$. 
Nearest neighbour 
sites in the tree correspond to nearest neighbour sites in the dual graph. The 
correspondence is not one-to-one, however, 
if $b$ and $c$ are nearest neighbour of the same
site of the tree, then they represent different sites of the dual, i.e. 
$\alpha(b)\not=\alpha(c)$. Consequently, calling 
$\tilde{z}_{a}$ the coordination number of a site of the tree and $z_{\alpha}$ 
the coordination of a site of the dual, we have 
$\tilde{z}_{\alpha}\leq z_{\alpha(a)}$. One site of the tree is called 
the origin $o$. We have $\alpha(o)=\alpha'$ and the maximum distance of 
any sites of the tree from $o$ is $k$ ($k$ will be called the length of the 
tree). There are exactly $N$ sites at a distance of $k$ steps 
from $o$ and to each of them we associate, in a one-to-one
correspondence, one of the loops. These sites will be called ending points $e$. 
$\alpha(e)=\alpha'$
and they are the only ones in the tree with coordination $1$. Therefore, the 
number of loops represented in the tree can be evaluated by counting the ending 
points. Finally, the sites crossed in the tree by the walk of $k$ steps, 
connecting the origin to an ending point $e$, correspond, in a one-to-one 
relation, to the sites of the dual graph belonging to the loop associated to 
$e$.

The set made of any single loop can be 
represented by a linear chain of $k+1$ sites and $k$ links.
The first site of the chain is the origin and it corresponds to $\alpha'$, the 
second corresponds to the second site of the loop and so on, the last site 
corresponds again to $\alpha'$ and it is the ending point of the tree. 

Let us show that, if there is a tree representation for any set of $N-1$ loops 
starting from $\alpha'$, we can add to any of these sets a new loop and produce 
a tree representing this new set. We will proceed by adding a new chain to
the tree, relevant to the $N-1$ loops. The sites of this tree correspond,
in both representations, to the same sites of the dual graph and they represent
the first $N-1$ loops. The starting point $\alpha'$ of the new loop is naturally
associated to the origin $o$ of the tree. Then, if one 
of the nearest neighbour of $o$ corresponds in the dual 
to the second site of the loop, we move to this site
of the tree and check if one of its neighbours correspond to the third site of
the loop and so on. Until we find that, at the step $q$, no neighbour of the 
site represents the site reached in the $q+1$ step of the loop. Since 
we are considering SOS loops (we can not move backward), this site 
is distant $q$ from the origin of the tree. Now we add to this site a 
bifurcation and a new chain of length $k-q$. The first site of this new chain 
corresponds to the site reached in the $q+1$ step of the loop, 
the second to the site reached in the $q+2$ step, the last corresponds again to 
$\alpha'$. So, when we add a loop to the set, we add to the tree one
ending point naturally 
associated with the new loop. The new neighbour sites added to the tree are 
neighbour also in the dual. Furthermore, we never add to any site of the tree 
$a$ a new neighbour $b$, such that $\alpha(b)$ is already 
represented in the tree by a neighbour of $a$.
This way, we produce a tree satisfying TP properties and representing a set of 
$N$ loops passing $\alpha'$. 

In particular, it is possible to construct the tree representing 
the set of all SOS loops of length $k$ surrounding $i$ and 
starting from $\alpha$. The number of these loops is equal to the number of the
ending points in the tree. In Figure 6 is illustrated
a tree corresponding to the all the loops of length $7$ passing $\alpha=0$ and
enclosing $i$.

\begin{figure}
\begin{center}
\leavevmode
\hbox{%
\epsfysize=6cm
\epsfbox{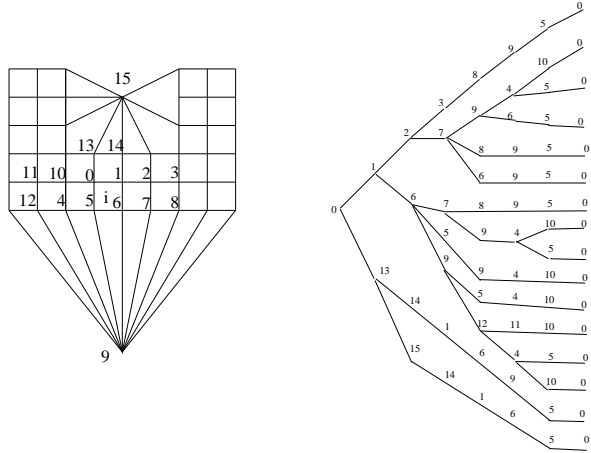}}
\caption{The tree corresponding to all loops of length $7$ 
passing $\alpha=0$ and enclosing $i$ (the loop are represented in the tree
in the clockwise direction).}
\end{center}
\end{figure}

Let us consider a tree of length $k$ and satisfying TP. We 
call $\bar{e}$ the 
ending point maximizing the product $\bar{P}_k$ of the 
coordination numbers  
of the sites crossed by the $k$ steps walk connecting the 
origin to $\bar{e}$. We prove by induction on $k$ that the number of 
the ending points $T_k$ is smaller than $\bar{P}_k$. 

For $k=1$ the tree is given by the origin and its $\tilde{z}_o$ neighbours 
which are the ending points. Hence both $\bar{P}_1$ and $T_1$ are equal to
$\tilde{z}_o$. Let us now suppose that $T_{k-1}\leq\bar{P}_{k-1}$ 
for any tree of length $k-1$. We need to show that, given a tree of length $k$, 
$T_k\leq\bar{P}_{k}$. Cutting the origin $o$ from the tree of length $k$, the 
graph is disconnected into $\tilde{z}_o$ subgraphs. Each of these subgraphs is 
a tree of length $k-1$, satisfying TP, whose origin is one of the 
neighbours of $o$. We will denote these trees with the index $h$ and we will 
call $T^h_{k-1}$ the number of the ending points of the tree $h$. $T_k$
is equal to the sum over $h$ of $T^h_{k-1}$. Hence:
\begin{equation}
\label{eq9}
T_k=\sum_{h=1}^{\tilde{z}_o} T^h_{k-1}
\leq \sum_{h=1}^{\tilde{z}_o} \bar{P}^h_{k-1}\leq \tilde{z}_o
\max_h(\bar{P}^h_{k-1})
\end{equation} 
In the first inequality we use the fact that $T^h_{k-1}$ and 
$\bar{P}^h_{k-1}$ are, respectively, the number of
the ending points and the maximum of the products of the coordination numbers 
in a tree of length $k-1$. Since the origin belongs to any of the walks
connecting $o$ to an ending point, $\bar{P}_k$ can be expressed as the
product of $z_o$ and the maximum value of $\bar{P}^h_{k-1}$. Therefore, from 
(\ref{eq9}) one immediately obtains the proof.

In the tree representing all the SOS loops of length $|l|$ enclosing $i$ and 
passing $\alpha$, we will call 
$\bar{w}_i^{\alpha}$ the walk connecting the origin to $\bar{e}$ and 
we will call $\bar{l}_i^{\alpha}$ the loop associated to $\bar{e}$. We have:
\begin{equation}
\label{eq10}
n_i^{\alpha}(|l|)\leq \prod_{a \in \bar{w}_i^{\alpha}}\tilde{z}_{a}
=\tilde{Z}_{\bar{l}_i^{\alpha}} 
\prod_{a \in \bar{w'}_i^{\alpha}}\tilde{z}_{a}
\leq \tilde{Z}_{\bar{l}_i^{\alpha}} 
\exp\left(\sum_{\gamma\in \bar{l'}_i^{\alpha}}\log(z_{\gamma})\right)
\end{equation} 
$\tilde{Z}_{\bar{l}_i^{\alpha}}$ is the coordination number of the tree site
$m$ corresponding to the site $\mu$ of the loop $\bar{l}_i^{\alpha}$ with the 
highest coordination in the dual graph (if the sites of
highest coordination are more than one, we arbitrarily chose one of them). 
$\bar{w'}_i^{\alpha}$ and $\bar{l'}_i^{\alpha}$
are obtained by subtracting from the walk $\bar{w}_i^{\alpha}$ and the loop 
$\bar{l}_i^{\alpha}$, respectively, $m$ and $\mu$.
In the last inequality we use the one-to-one correspondence between 
$\bar{w'}_i^{\alpha}$ and $\bar{l'}_i^{\alpha}$ and fact that the coordination
number $\tilde{z}_{a}$ of the sites of the tree is smaller or equal than the
coordination number $z_{\gamma(a)}$ of the relevant sites of the dual graph.

\section{Sparse high connectivity}

In the dual graph the possible values of the coordination numbers are $4
\cdot 3^m$ with $m=0,1,\dots$. Hence we have
\begin{equation}
\label{eq11}
n_i^{\alpha}(|l|) \leq 
\tilde{Z}_{\bar{l}_i^p} 
\exp\left(\sum_{m} N_{\bar{l'}_i^{\alpha}}(m)(\log(4)+m\log(3))\right)
\end{equation} 
$N_{\bar{l'}_i^{\alpha}}(m)$ is the number of sites in $\bar{l'}_i^{\alpha}$ 
with coordination number in the dual graph equal to $4\cdot 3^m$. 

The minimum chemical distance $d(m)$ between two sites of the dual graph
with coordination greater or equal than $4\cdot 3^m$ is:
\begin{equation}
\label{eq12}
d(m)= \left\{
\begin{array}{cl}
1 & {\rm if } \ m=0 \cr
2^{m+1} & {\rm if } \ m>0 \cr
\end{array}
\right .
\end{equation} 
Equation (\ref{eq12}) represents the fundamental topological property of the
dual graph of the Sierpinski carpet which will allow us to prove the 
exponential bound on $n_{i}(k)$, even in absence of bounded coordination.
We will call this property sparse high connectivity, since it implies that, 
in the topology of the dual graph generated by the chemical distance, sites
with high coordination are far apart.

For a SOS loop $l$, the sites with 
coordination $\geq 4\cdot 3^{m}$ are less than $1+|l|/d(m)$. Since in 
$\bar{l'}_i^{\alpha}$ we have excluded the point of highest coordination, we can
overestimate the number of sites of coordination $4\cdot 3^{m}$ as: 
\begin{equation}
\label{eq13}
N_{\bar{l'}_i^{\alpha}}(m)\leq{|l|\over d(m)}
\end{equation} 
From (\ref{eq13}) we obtain that, if $m>\log_2(l)-1=m_M$, then 
$N_{\bar{l'}_i^{\alpha}}(m)=0$. Therefore, $m_M$ represents the maximum value 
of $m$ on the sites of $\bar{l'}_i^{\alpha}$. 

Thus we have:
\begin{eqnarray}
n_i^{\alpha}(l) & \leq &
\tilde{Z}_{\bar{l}_i^{\alpha}} 
\exp\left(|l|\log(4)+l\sum_m^{m_M}2^{-m-1}(\log(4)+m\log(3))\right)\nonumber\\
& \leq &
\tilde{Z}_{\bar{l}_i^{\alpha}} 
\exp\left(|l|\log(4)+l\sum_{m}^{\infty}2^{-m-1}(\log(4)+m\log(3))\right)\nonumber\\
& \leq &\tilde{Z}_{\bar{l}_i^{\alpha}} 
\exp\left(K|l|\right)
\label{eq14}
\end{eqnarray}
where $K$ is a suitable positive, real constant.

\section{The site of highest coordination}

In this section we will show 
that $\tilde{Z}_{\bar{l}_i^{\alpha}}\leq 10 l^{\log_2(3)}$.
If $z_{\mu}\leq 10 l^{\log_2(3)}$ ($\mu$ is the site in the dual graph with 
highest coordination), the inequality is trivially satisfied. 
Otherwise, we will use the fact that $\tilde{Z}_{\bar{l}_i^{\alpha}}$ 
is the coordination of a site of a tree
representing SOS loops surrounding $i$. Two sites of the
tree are neighbour only if the corresponding sites of the dual are 
neighbours and it exists a SOS loop passing through them and enclosing $i$.
Therefore, the bound on $\tilde{Z}_{\bar{l}_i^{\alpha}}$ is proved 
by showing that the ${Z}_{\mu,\bar{l}_i}\leq  10 l^{\log_2(3)}$; where 
${Z}_{\mu,\bar{l}_i}$ represents the number of nearest neighbours of $\mu$, 
which can be crossed by a SOS walk of length $|l|$, passing $\mu$ and enclosing 
$i$.

Let $\nu$ and $\rho$ be nearest neighbour of
$\mu$. We define the distance  along the boundary $\bar{r}_{\nu,\rho}$
as in Figure 7. 
In Figure 7 the plaquette of the site $\mu$ is also defined.
\begin{figure}
\begin{center}
\leavevmode
\hbox{%
\epsfysize=6cm
\epsfbox{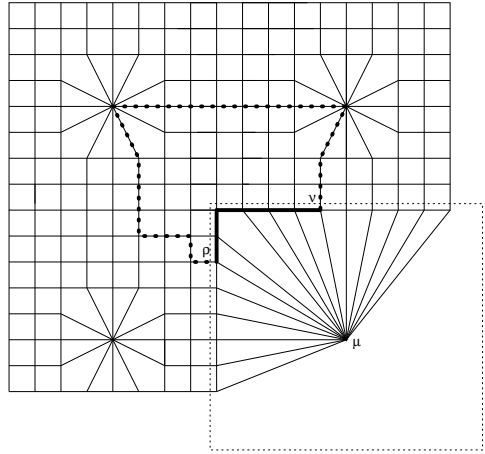}}
\caption{The plaquette of the site $\mu$ is given by the region denoted by the
dashed line. The sites within this region (excluded $\mu$) are the sites
belonging to the plaquette. The distance along the boundary between the sites of
the plaquette $\nu$ and $\rho$ is the length of the line plotted with the heavy
line. In this example $\bar{r}_{\rho,\nu}=6$. With the dotted lines we denoted a
walk in the dual graph between $\nu$ and $\rho$ of length 18.}
\end{center} 
\end{figure}

Let us consider a SOS loop $l^{\nu,\rho}$ of length $|l|$ passing $\mu$ and its
nearest neighbours $\rho$ and $\nu$.
We will prove that:
\begin{equation}
\bar{r}_{\nu,\rho}\leq 2|l|^{\log_2(3)}. 
\label{eq15}
\end{equation}
Furthermore, if $l^{\nu,\rho}$ encloses the site of the carpet $i$,  
from the planarity of the dual graph we have the following property:
no loop of length $|l|$ passing $\mu$ and enclosing $i$ can
pass through any site $\tau$ of the plaquette of $\mu$, such that $\tau$ is 
located with respect to $\rho$ in the opposite direction of $\nu$ at a 
distance $\bar{r}_{\tau,\rho}>4\cdot l^{\log_2(3)}$. 

An analogous 
property holds for the sites of the plaquette at a distance greater than 
$4\cdot l^{\log_2(3)}$ from $\nu$ in the direction opposite to $\rho$. 
Therefore, since $\bar{r}_{\nu,\rho}\leq 2|l|^{\log_2(3)}$, 
the number ${Z}_{\mu,\bar{l}_i}$ of sites belonging to the plaquette of
$\mu$ and crossed by a loop passing $\mu$ and surrounding $i$ is 
smaller than $10 |l|^{\log_2(3)}$. This way
we obtained the bound on $\tilde{Z}_{\bar{l}_i^{\alpha}}$.

Let us prove (\ref{eq15}). The dual graph is naturally embedded in a two
dimensional euclidean space. Since the plaquettes are convex subsets of this 
space, the length of the border $\bar{r}_{\nu,\rho}$ turns out to be 
smaller than the length, in the euclidean space, of any walk from $\nu$ to 
$\rho$ which do not pass through the plaquette itself. 
In particular, it is smaller than the length of 
$l'^{\nu,\rho}$. $l'^{\nu,\rho}$ is the walk obtained subtracting $\mu$ 
from the loop $l^{\nu,\rho}$. 
Now, a step in the dual graph, crossing a site with coordination $z_\rho$,
covers, in the euclidean space, a distance smaller than 
$z_{\rho}/2$. We have:
\begin{equation}
\bar{r}_{\nu,\rho}  \leq 
{1\over 2}\sum_{\rho\in l'^{\nu,\rho}} z_\rho 
\leq 2\sum_{m}^{m_M} N_{l'^{\nu,\rho}}(m) 3^m 
\leq 2l+ l \sum_{m}^{\log_2(|l|)-1} \left({3\over 2 }\right)^m
\leq  2l^{\log_2(3)}
\label{eq16}
\end{equation}
$N_{l'^{\nu,\rho}}(m)$ is the number of sites in $l'^{\nu,\rho}$ with
coordination $4\cdot 3^m$. In equation (\ref{eq16}) $N_{l^{\nu,\rho}}(m)$ has 
been estimated using inequality (\ref{eq13}).

Now we need to prove that no that no loop of length $|l|$ passing $\tau$ and 
$\mu$ can enclose the site $i$. We suppose, ad absurdum, the existence of 
this loop and we call it $l^{\tau,\sigma}$ ($\sigma$ is the other site of the 
loop belonging to the plaquette of $\mu$). Since $\tau$ and $\sigma$ are 
connected by a loop of length $|l|$, 
from (\ref{eq15}) we have $\bar{r}_{\tau,\sigma}<2|l|^{\log_2(3)}$.
Hence $\bar{r}_{\rho,\sigma}>2|l|^{\log_2(3)}$.
If also $l^{\tau,\sigma}$ surrounded $i$, for the planarity of the dual graph,
there should exist two sites $\gamma$ and $\delta$ belonging both to 
$l^{\tau,\sigma}$ and to $l^{\nu,\rho}$ (see Figure 8 for an illustration).

\begin{figure}
\begin{center}
\leavevmode
\hbox{%
\epsfysize=6cm
\epsfbox{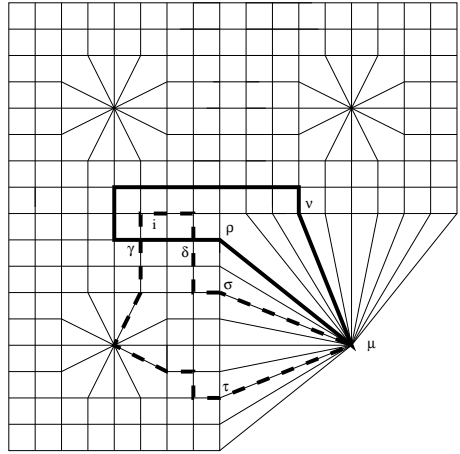}}
\caption{With the heavy continuous line we denoted $l^{\nu,\rho}$, with the
dashed line $l^{\tau,\sigma}$.}
\end{center}
\end{figure}

We denote with $|\gamma,\delta|_{l^{\tau,\sigma}}$ the length of the 
part of the walk $l'^{\tau,\sigma}$ within the sites $\gamma$ and $\delta$. 
We have
\begin{eqnarray}
|\rho,\delta|_{l^{\nu,\rho}}+|\delta,\gamma|_{l^{\nu,\rho}}+|\gamma,\nu|_{l^{\nu,\rho}}=|l|-2\nonumber\\
|\sigma,\delta|_{l^{\tau,\sigma}}+|\delta,\gamma|_{l^{\tau,\sigma}}+|\gamma,\tau|_{l^{\tau,\sigma}}=|l|-2
\label{eq17}
\end{eqnarray}
and then:
\begin{equation}
|\rho,\delta|_{l^{\nu,\rho}}+|\gamma,\nu|_{l^{\nu,\rho}}+|\sigma,\delta|_{l^{\tau,\sigma}}+|\gamma,\tau|_{l^{\tau,\sigma}}
\leq 2(|l|-2)
\label{eq18}
\end{equation}
On the other hand, since $\bar{r}_{\nu,\tau}>|l|^{\log_2(3)}$ and 
$\bar{r}_{\rho,\sigma}>|l|^{\log_2(3)}$, from (\ref{eq15}) we get
\begin{eqnarray}
|\rho,\delta|_{l^{\nu,\rho}}+|\sigma,\delta|_{l^{\tau,\sigma}} > |l|-2\nonumber\\
|\nu,\gamma|_{l^{\nu,\rho}}+|\tau,\gamma|_{l^{\tau,\sigma}} > |l|-2
\label{eq19}
\end{eqnarray}
and then
\begin{equation}
|\rho,\delta|_{l^{\nu,\rho}}+|\sigma,\delta|_{l^{\tau,\sigma}}+|\nu,\gamma|_{l^{\nu,\rho}}+|\tau,\gamma|_{l^{\tau,\sigma}} 
>2(|l|-2) 
\label{eq20}
\end{equation}
Equation (\ref{eq20}) can not hold together with equation (\ref{eq18}). 
Therefore, assuming the existence of the site $\tau$, we obtained the 
contradiction we wanted to prove. Hence, we have concluded 
the proof of the bound on $\tilde{Z}_{\bar{l}_i^{\alpha}}$.

\section{Euclidean length of a loop}

With the bound on $\tilde{Z}_{\bar{l}_i^{\alpha}}$ proved in the previous
section, we obtain that probability for the site $i$ to have negative
magnetization can be estimated as
\begin{equation}
\label{eq21}
p_i^-  \leq \sum_{|l|=0}^{\infty} N_i(|l|) 10 |l|^{\log_2(3)} e^{(K-\beta) |l|}
\end{equation} 
$N_i(|l|)$ is the number of points in the dual lattice which can be crossed
by a SOS of length $|l|$ which encloses $i$. 

An estimate of $N_i(|l|)$ can be obtained by using again the fact that the dual
graph is embeddable in a two-dimensional euclidean space. In particular, in a 
walk in the dual graph, a step crossing the site $\alpha$ covers in the 
euclidean space a distance smaller or equal than
$z'_{\alpha}=\bar{r}_{\alpha^-,\alpha^+}$, where $\alpha^-$ and $\alpha^+$ are
respectively the sites preceding and following $\alpha$ within the walk.
Let us consider $l_i$ SOS loop in the dual graph of length $|l|$ and enclosing 
the site $i$ of the carpet. We call $d(l_i)$ the distance covered by this loop 
in the euclidean space, we get
\begin{equation}
\label{eq22}
d(l_i)\leq \sum_{\alpha\in l_i} z'_{\alpha}
\leq Z'_{l_i} +\sum_{\alpha \in l'_i} z_{\alpha}
\end{equation} 
In the sum we separate the contribution of the site with highest
coordination number and we used the simple property that 
$z'_{\alpha}<z_{\alpha}$. 
$Z'_{\bar{l}_i}$ represents the distance along the boundary between the sites
preceding and following the site of highest coordination number in a walk of
length $|l|$. In the previous section we proved that $Z'_{\bar{l}_i}\leq 2 
l^{\log_2(3)}$. Furthermore the sum over all sites belonging to $l'_i$ has been
already evaluated in (\ref{eq16}). Therefore we have that
\begin{equation}
\label{eq23}
d(l_i)\leq 8 |l|^{\log_2(3)}
\end{equation} 
The number of sites of the dual graph, distant from $i$ in
the euclidean space less than $d(l_i)$ it is smaller than $\pi d^2(l_i)$. 
Therefore we obtain that $N_i(l)\leq \pi (8|l|^{\log_2(3)})^2$. Hence
\begin{equation}
\label{eq24}
p_i^-  \leq \sum_{|l|=0}^{\infty} C |l|^{3\log_2(3)} e^{(K-\beta) |l|}
\end{equation} 
where $C$ is a suitable real constant.
For small enough values $\beta$ the sum in (\ref{eq24}) is smaller than 
$\epsilon<1/2$. Thus we obtain the uniform bound on $p_i^-$ proving the
existence of spontaneous magnetization at finite temperature.

\ack

The author is grateful to R. Burioni and D. Cassi for useful 
comments and discussions.


\section*{References}
\begin{thebibliography}{13}
\bibitem{mandel}
Mandelbrot BB {\it The fractal geometry of Nature} (Freeman: San Francisco)
\bibitem{rgising}
Gefen Y, Aharony A and Mandelbrot BB 1984 {\it J. Phys. A} {\bf 17} 1277 \nonum
Wu YK and Hu B 1987 {\it Phys. Rev. A} {\bf 35} 1404\nonum
Bonnier B, Leroyer Y and Meyers C 1988 {\it Phys. Rev. B} {\bf 37} 5205
\bibitem{rgcarpet}
Hattori K, Hattori T and Watanabe H 1985 {\it Phys. Rev. A} {\bf 32} 3730\nonum
Lai P and Goldschmit 1987 {\it J. Phys. A} {\bf 20} 2159\nonum
Chame A and Costa UMS 1990 {\it J. Phys. A} {\bf 23} L1127\nonum
Kim MH, Yoon DH and Kim I 1993 {\it J. Phys. A} {\bf 26} 5655
\bibitem{htexpising}
Bonnier B, Leroyer Y and Meyers C 1989 {\it Phys. Rev. B} {\bf 40} 8961\nonum
Fabio DA, Ar\~ao Reis A and Riera R 1994 {\it Phys. Rev. E} {\bf 49} 2579
\bibitem{numising}
Banhot G, Neuberger H and and Shapiro JA 1984 {\it Phys. Rev. Lett.} 
{\bf 58} 6386\nonum
Angles D'Auriac JC and Rammal R 1986 {\it J. Phys. A} {\bf 19} L655\nonum
Monceanu P, Perreau M and Heb\'ert F 1998 {\it Phys. Rev. B} {\bf 58} 6386\nonum
Zheng GP and Li M 2000 {\it Phys. Rev. E} {\bf 62} 6253\nonum
Pruessner G, Loison D, and Schotte KD 2001 {\it Phys. Rev. B} {\bf 64} 134414
\bibitem{numcarpet}
Dasgupta R, Ballabh TK and Tarafdar S 1999 {\it J. Phys. A} {\bf 32} 6503
\bibitem{gasketising}
Gefen Y, Aharony A and Mandelbrot BB 1984 {\it J. Phys. A} {\bf 17} 435 \nonum
Higuchi Y and Yoshida N 1996 {\it J. Stat. Phys.} {\bf 84} 295
\bibitem{gasket}
Rammal R 1984 {\it J. Phys. (Paris)} {\bf 45} 191\nonum
Stinchcombe RB 1990 {\it Phys. Rev. B} {\bf 41} 2510
\bibitem{barlow}
Barlow MT and Bass RF 1999 in {\it Random Walks and Discrete Potential Theory}
ed M Piccardello and W Woess (Cambridge: University Press) 26
\bibitem{specdim}
Alexander S and Orbach R 1987 {\it J. Phys. (Paris) Lett.} {\bf 92} {108}\nonum
Hattori K, Hattori T and Watanabe H 1987 {\it Prog. Theor. Phys. Suppl.} 
{\bf 92} {108} \nonum
Burioni R and Cassi D 1997 {\it Mod. Phys. Lett.} {\bf B 11}S {1095}  
\bibitem{mwgen}
Cassi D 1996 {\it Phys. Rev. Lett.} {\bf 76} 2941
\bibitem{shinoda}
Shinoda M 2002 {\it J. App. Prob.} {\bf 39} 1
\bibitem{peierls}
Peierls R 1936 {\it Proc. Camb. Philos. Soc.} {\bf 32}  477
\bibitem{peierls2}
Griffiths RB 1964 {\it Phys. Rev.} {\bf 136} 437\nonum
Huang K 1987 {\it Statistical mechanics} (New York: John Wiley and Sons, Inc.)
\bibitem{chayes}
Chayes JT, Chayes L and Fr\"ohlich J 1985 {\it Comm. Math. Phys.} {\bf 100} 399
\end{thebibliography}
\end{document}